\documentclass[aps,prl,letterpaper,floatfix,twocolumn,groupedaddress]{revtex4}
\usepackage{graphicx}

\newcommand{\be}{\begin {equation}}
\newcommand{\ee}{\end {equation}}
\newcommand{\bea}{\begin {eqnarray}}
\newcommand{\eea}{\end {eqnarray}}

\begin{document}


\title{On the origin of  non-Gaussian statistics in hydrodynamic turbulence}



\author{Charles Meneveau}
\affiliation{Department of Mechanical Engineering and Center for
Environmental and Applied Fluid Mechanics, The Johns Hopkins
University, Baltimore, MD 21218}
\author{Yi Li}
\affiliation{Department of Mechanical Engineering and Center for
Environmental and Applied Fluid Mechanics, The Johns Hopkins
University, Baltimore, MD 21218}


\date{\today}

\begin{abstract}
Turbulent flows are notoriously difficult to
describe and understand based on first principles. One reason is
that  turbulence contains highly intermittent bursts of vorticity
and strain-rate
with highly
non-Gaussian statistics. Quantitatively, intermittency is manifested in highly elongated
tails in the probability density functions of the velocity
increments between pairs of points.  A long-standing open  issue
has been to predict the origins of intermittency and non-Gaussian
statistics  from the Navier-Stokes equations.  Here we derive,
from the Navier-Stokes equations, a simple nonlinear dynamical
system for  the Lagrangian evolution of
longitudinal and transverse velocity increments.  From this system
we are able to show  that the ubiquitous non-Gaussian tails in
turbulence have their origin in the inherent  self-amplification
of longitudinal velocity
increments, and cross amplification of  the transverse velocity increments.
\end{abstract}

\pacs{}

\maketitle

Intermittency in turbulent flows refers to the violent and extreme
bursts of vorticity and rates of strain that occur interspersed
within regions of relatively quiet flow\cite{Frisch95}. These
infrequent, but extreme events are believed to cause observed
deviations from the classical Kolmogorov theory of
turbulence\cite{Kolmogorov41}. Intermittency also has a number of
practical consequences since it can lead to sudden emergence of
strong vortices in geophysical flows\cite{Sreeni99},  to
modifications of the local propagation speed of turbulent flames
\cite{Peters99}, etc.  One of the observable manifestations of
intermittency is the tendency of velocity increments, i.e. the
difference between velocities at two spatial points separated by a
distance $\ell$, to display highly non-Gaussian statistics when
$\ell$ is smaller than the flow integral scale, $L$. The tails of
velocity-increment probability density functions (pdf) are
observed to be exponential and even stretched
exponential\cite{Frisch95,Sreeni99}. Moreover, an inherent
asymmetry develops in the distribution of the longitudinal
velocity increments, i.e. the difference of the velocity component
in the direction of the displacement between the two points. This
asymmetry yields the well-known negative skewness of longitudinal
velocity increments\cite{Frisch95}. While the negativity of
skewness  can be derived from the Navier-Stokes (N-S) equations in
isotropic turbulence\cite{Kolmogorov41,Frisch95}, a
straightforward mechanistic explanation of the origins of
stretched exponential tails, intermittency, and asymmetry has
remained elusive.

In one dimension for the Burgers equation, the emergence of
negative skewness and long negative tail in the pdf starting from
random initial conditions is well understood based on the
nonlinear term's tendency to steepen the velocity gradient. In 3D
turbulent flows, the notion of nonlinear ``self-amplification'' as
the cause of intermittency has long been
suspected\cite{Zeffetal03}. Yet, these expectations have eluded
quantitative analysis due to the difficulty in deriving
lower-dimensional models that maintain the relevant information
about the vectorial nature of the full 3D dynamics. Many surrogate
models have been proposed, such as shell models \cite{Biferale03},
the mapping closure\cite{Kraichnan90,SheOrszag91}, etc, but the
connection with the original N-S equations is typically based on
qualitative and dimensional resemblances instead of on systematic
derivation.

\begin{figure}[h]
\centering
\includegraphics[width=0.7\linewidth]{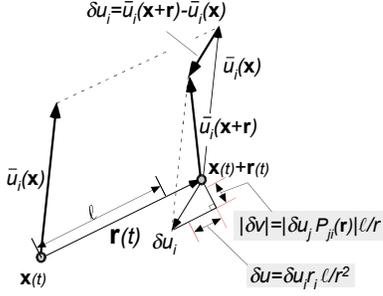}
\caption{Illustrative sketch of velocity increment $\delta
u_i({\bf r})$ between two points ${\bf x}(t)$ and ${\bf x}(t)+{\bf
r}(t)$, and the components of $\delta u_i({\bf r})$ longitudinal
and transverse to the displacement vector ${\bf r}$. The
quantities of interest are $\delta u$ and $\delta v$, defined as
the components of the velocity increment over a fixed length
$\ell$.}\label{fig:sketch}
\end{figure}

We consider the coarse-grained N-S equations filtered at scale
$\Delta$ comparable (and larger) than the scale $\ell$. Let
$\overline{u}_i$ be the filtered velocity field. Defining the
velocity gradient tensor $\overline{A}_{ji} = \partial
\overline{u}_i/\partial x_j$ and taking the gradient of the
filtered N-S equations one obtains\cite{Vieillefosse84,Cantwell92}
that the rate of change of the velocity gradient is given by
\be
\dot{\overline{A}}_{ji} =
-\left(\overline{A}_{jk}\overline{A}_{ki} + 2Q/3~
\delta_{ji}\right) + H_{ji},
\ee
where $Q=-\overline{A}_{mn}\overline{A}_{nm}/2$ arises from
continuity. The tensor $H_{ji}$ contains the trace-free part of
the pressure Hessian, subgrid, and viscous force
gradients\cite{BorueOrszag98,VanderBosetal02}: $
H_{ji}=-(\partial^2_{ji}\overline{p}-\frac{1}{3}\delta_{ij}\partial^2_{kk}\overline{p})
-(\partial^2_{jk}\tau_{ik}-\frac{1}{3}\delta_{ij}\partial^2_{lk}\tau_{lk})
+\nu\partial^2_{kk}\overline{A}_{ji}$, in which $\overline{p}$ is
the filtered pressure divided by density and $\nu$ the viscosity.
$\tau_{ij}=\overline{u_i u}_j -\overline{u}_i\overline{u}_j$ is
the subgrid-scale (SGS) stress. The time derivative $\dot{(\ )}$
is a Lagrangian material derivative defined as the rate of change
of the gradient tensor following the local smoothed flow. Setting
$H_{ij}=0$ yields the so-called ``Restricted Euler"
dynamics\cite{Vieillefosse84,Cantwell92}. A fruitful method to
model the effects of $H_{ij}$ has been to track material
deformations using either tetrad dynamics\cite{Chertkovetal99} or
the Cauchy-Green tensor\cite{JeongGirimaji03}. Here we focus on a
simpler object - a line element, aiming at identifying the
mechanism generating intermittency. Thus, consider two points
separated by a displacement vector $\bf r$ of length smaller than,
or of the order of, $\Delta$ so that the local velocity field is
smooth enough to be approximated as a linear field. The velocity
increment between the two points over the displacement $\bf r$ is
then
\be
\delta u_i({\bf r},t) \equiv \overline{u}_i({\bf x} + {\bf r}) -
\overline{u}_i({\bf x}) \approx \overline{A}_{ki}~ r_k .
\label{eq:defdu}
\ee
The longitudinal and transverse velocity increments, $\delta
u(r,t)$ and $\delta v(r,t)$ respectively, can be evaluated from
the two projections of velocity increment Eq. \ref{eq:defdu} into
directions longitudinal and transverse  to ${\bf r}$  (see FIG.
\ref{fig:sketch}):
\be
\delta u(r,t) = \overline{A}_{ki}~ r_k~\frac{r_i}{r}, ~~~~  \delta
v(r,t) = \left|P_{ij}({\bf r})\overline{A}_{kj}~ r_k \right|,
\label{eq:defdudv}
\ee
where $P_{ij}({\bf r})=\delta_{ij}-r_ir_j/r^2$ and $r=|\bf r|$.

Note that $\delta u(r,t)$ and $\delta v(r,t)$ correspond to
velocity increments over a displacement $r_i(t)$ that is evolving, in a
local linear flow, according to equation $\dot{r}_i=
\overline{A}_{mi}~ r_m$. To study the evolution of velocity
increments at a fixed scale $\ell$, it is necessary to eliminate
effects from the changing distance between the two points.
Consider a line that goes through the two points. Still within the
assumption of a locally linear velocity field, the velocity
increments across a fixed distance $\ell$ along this line are
$\delta u \equiv \delta u(r,t) {\ell}/{r}$, $\delta v \equiv
\delta v(r,t){\ell}/{r}$ (see FIG. \ref{fig:sketch}).

Taking time derivatives of $\delta u$ and $\delta v$, and using
the expressions for $\dot{\overline{A}}_{ji}$ and $\dot{r}_i$,
many terms simplify and one arrives at the following ``advected
delta-vee'' system of equations:
\bea
\delta \dot{u} &=&   -{ \delta u^2}~{\ell}^{-1} +{ \delta
v^2}~{\ell}^{-1} - \frac{2}{3} Q \ell + Y, \label{eq:dudt} \\
\delta \dot{v} & =&  - {2~\delta u~ \delta v}~{\ell}^{-1} + Z,
\label{eq:dvdt}
\eea
where $Y=\ell H_{ij}r_i r_j/r^2$ and $Z=\ell H_{ij} e_j r_i/r$
contain the anisotropic nonlocal effects of the pressure,
inter-scale effects of subgrid-scale stresses, and damping effects
of molecular viscosity (${\bf e}$ is a unit vector in the
direction of the transverse velocity component).  The first term
on the right-hand-side (rhs) of the equation for $\delta \dot{u}$
also occurs in 1D Burgers equation (the self-amplification effect
of negative velocity gradients). The second term indicates that
the transverse velocity (rotation) tends to counteract the
self-amplification process.  For $\delta \dot{v}$, the first term
on rhs of Eq. \ref{eq:dvdt} suggests exponential growth of $\delta v$ at a rate $|\delta u |$ when $\delta u <
0$. This ``cross-amplification'' mechanism can lead to very large
values of $|\delta v|$.

We now pose the question whether the growth of intermittency and
the asymmetry of longitudinal velocity increments can be
understood based on this system of equations, but without the
 effects represented by
$Y$ and $Z$ (i.e., ``Restricted Euler" dynamics). In order to
determine whether this simplified system approximates $\delta
\dot{u}$ and $\delta \dot{v}$ in real turbulence, comparisons are
made with direct numerical simulations (DNS). The rates of change
of $\delta u$ and $\delta v$ predicted by DNS are obtained by
finite difference in time from two DNS velocity fields separated
by the simulation time-step $\delta t=0.001$.  The data are
obtained from a  pseudo-spectral simulation of the N-S equations,
with $256^3$ nodes and Taylor-scale Reynolds number $R_\lambda
\approx 162$. The velocity fields are coarse-grained using a
Gaussian filter of characteristic length $\Delta = 40\eta$, where
$\eta$ is the Kolmogorov length scale, yielding filtered velocity
fields $\overline{u}_i({\bf x},t_0)$ and $\overline{u}_i({\bf
x},t_0+\delta t)$  ($i=1, 2, 3$). At the initial time $t_0$, to
every grid-point ${\bf x}(t_0)$ on the computational mesh, we
associate a partner ${\bf x}(t_0)+{\bf r}(t_0)$ at a distance
$|{\bf r}(t_0)|=\ell = 40 \eta$ in some Cartesian direction. For
each pair of points we measure the longitudinal and transverse
velocity increments. Then, we find the position to which ${\bf
x}(t_0)$ and ${\bf x}(t_0)+{\bf r}(t_0)$ will be advected by the
smoothed velocity field, which are, using simple Euler
integration,
 ${\bf x}(t_0+\delta t)={\bf x}(t_0)+\overline{\bf u}({\bf x},t_0)
\delta t$, and ${\bf x}(t_0+\delta t)+{\bf r}(t_0+\delta t)$,
where ${\bf r}(t_0+\delta t)={\bf r}(t_0)+[\overline{\bf u}({\bf
x}(t_0)+{\bf r}(t_0),t_0)-\overline{\bf u}({\bf
x}(t_0),t_0)]\delta t$ is the new displacement vector.  The final
end-point at a fixed distance $\ell$ is found by moving the
material end-point  ${\bf x}(t_0+\delta t)+{\bf r}(t_0+\delta t)$
along the new displacement vector to the point ${\bf x}(t_0+\delta
t)+{\bf r}(t_0+\delta t)\ell/|{\bf r}(t_0+\delta t)|$, so that the
distance is kept fixed. Velocities at the new locations are
obtained from the stored field at the new time using bilinear
interpolation, and the longitudinal and transverse components are
evaluated, by projections onto direction parallel and
perpendicular to the new displacement vector between the two
points. The rate of change of $\delta u$ and $\delta v$ is
evaluated using first-order finite difference in time.
Conversely, the rates of change
predicted by the model system are evaluated as $-\delta u^2/\ell +
\delta v^2/\ell - (2/3) Q \ell$ and $-2~\delta u ~\delta v/\ell$
from the measured values of $\delta u$, $\delta v$ and $Q$.  Both
real and modeled rates of change are computed over a large number
of points in the DNS data, and their correlation coefficient and
joint  pdf are evaluated.
\begin{figure}[ht]
\begin{minipage}[h]{0.5\linewidth}
\includegraphics[width=\linewidth]{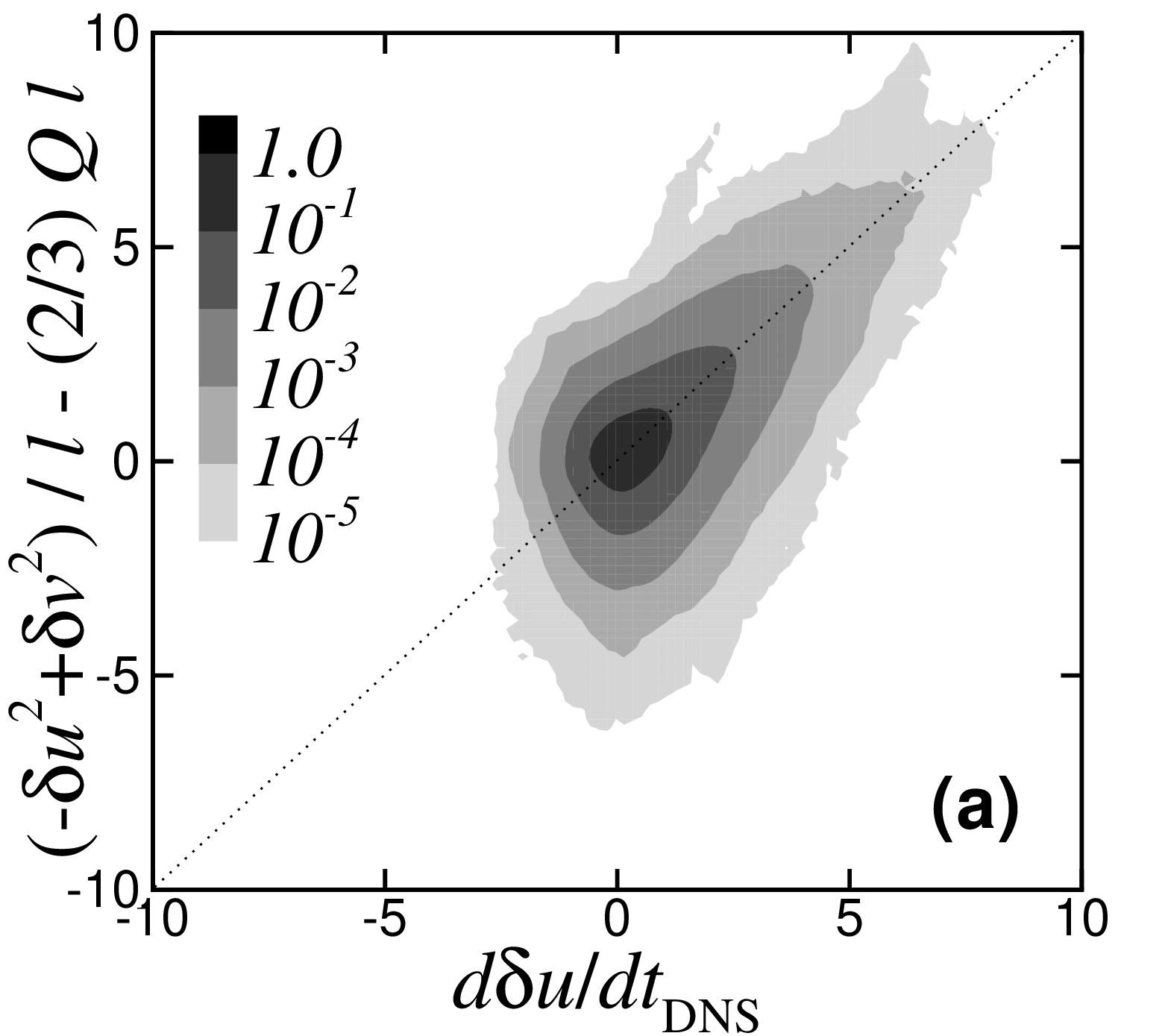}
\end{minipage}%
\begin{minipage}[h]{0.5\linewidth}
\includegraphics[width=\linewidth]{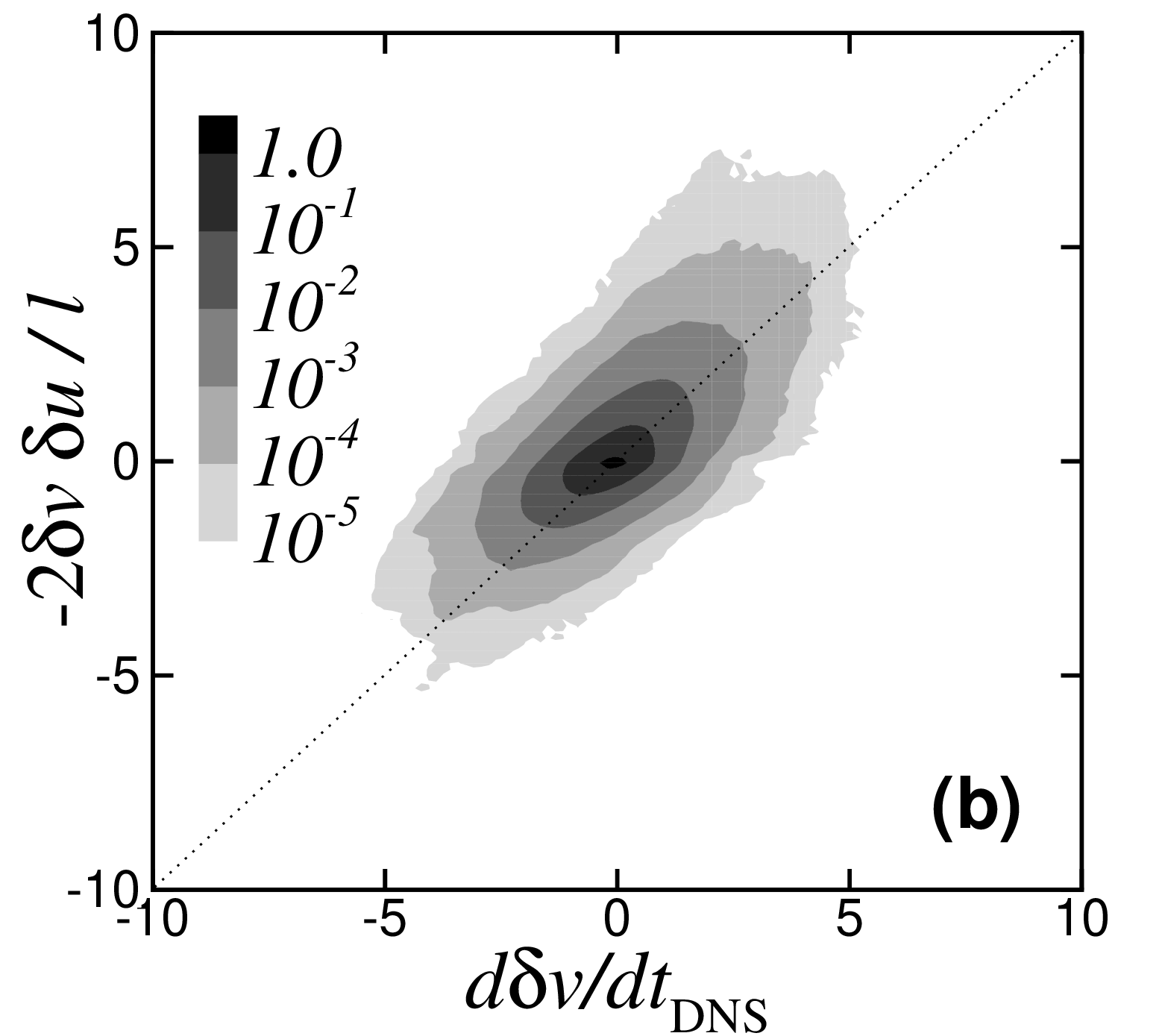}
\end{minipage}
\caption{Joint pdf of rates of change of velocity increments
predicted from DNS (filtered at $\Delta=40 \eta$ and taking
increments over a distance $\ell=40\eta$) and the ``advected
delta-vee" system. (a) longitudinal and (b) transverse velocity
increments. Results are robust with changes in $\Delta$ and $\ell$
(with $\ell \leq \Delta$).} \label{fig:jpdf}
\end{figure}

FIG. \ref{fig:jpdf} shows the joint pdf of the model results
versus the rates of change measured from DNS. A clear correlation
can be seen between model and DNS results. Correlation
coefficients are $0.54$ for the longitudinal and $0.61$ for the
transverse velocity increments, indicating that the model system
captures important (but clearly not all) effects seen in the real
dynamics. The deviations between model system and DNS are caused
by the neglected $Y$ and $Z$ terms, to be studied in future work.

\begin{figure}[ht]
\centering
\includegraphics[width=0.8\linewidth]{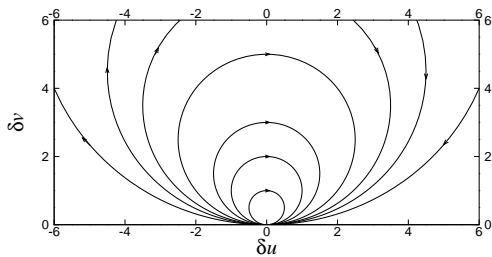}
\caption{Phase-space portrait of the ``advected delta-vee''
dynamical system ${\delta \dot{u}} = -\delta u^2+\delta v^2$,
$\delta \dot{v} = -2 ~\delta u ~\delta v$ (for $Q_0=0$ and
$\ell=1$).}\label{fig:phase}
\end{figure}

After confirming that the simplified system captures important
trends in 3D fluid turbulence, we explore the trends predicted by
solutions of the model system. In the present Letter we set $Q$ to a
constant $Q_0$ (numerical tests show that allowing $Q$ to evolve
in time leads to the same short-time behavior to be displayed
below, except if $Q$ were to be closely correlated with $\delta u$
and $\delta v$, which is not the case in 3D turbulence, since $Q$
depends on velocity gradients along two additional
directions\footnote{\label{2Dapp} In 2D turbulence, to which the
model can also be applied, the term $-2Q\ell/3$ must be replaced
with $-\ell \det(\bf{\overline{A}})$. It can be shown that one of
the two terms of the determinant exactly cancels the term $-\delta
u^2$ in the equation for $\delta u$. This cancels the mechanism
for growth of negative skewness and intermittency in 2D. In 3D
there is no full cancellation due to the weaker correlations
among the different directions. More detailed results for the 2D
case will be reported elsewhere.}.). We note that for $Q=0$, the
system describes the relative motion of a fluid (with a locally
linear velocity field) consisting of non-interacting
(``ballistic'') particles that maintain their initial velocity.
For $Q \neq 0$, the particles are subjected to a relative force
equal to the spherical average of the pressure, inter-scale, and
viscous damping forces.  For the case $Q_0=0$ the analytical
solution  is
\bea
& &\delta u(t)=\ell
e_0[e_0t+\delta u_0 \ell]/\{[e_0 t + \delta u_0 \ell]^2+\delta
v_0^2 \ell^2\},\\
& &\delta v(t)=\ell^2 \delta v_0 e_0/\{[e_0 t +
\delta u_0 \ell]^2+\delta v_0^2 \ell^2\},
\eea
where $e_0=\delta
u_0^2+\delta v_0^2$. For discrete values of time, this defines a
mapping (the ``advected delta-vee map''). The system has an
invariant
\be
U_0= (\delta u^2+\delta v^2)/\delta v,
\ee
and its
(circular) phase-space trajectories are $\delta u^2 + (\delta
v-U_0/2)^2=(U_0/2)^2$, as shown in FIG. \ref{fig:phase}.

In order to illustrate the evolution of $\delta u(t)$ and $\delta v(t)$,
we start from an ensemble of randomly oriented lines for which the velocity increment vectors are
initialized from a Gaussian distribution.
The increments $\delta u(t)$ and $\delta v(t)$ over these lines are evaluated at
several later times. To compare with experimental data, two issues need to be considered.
First, since $\delta v(t)$ is the magnitude of the transverse velocity increment vector, it has to be projected
onto a coordinate direction to obtain a component of the transverse increment, $\delta v_c =\delta v \cos\theta$.
For isotropic turbulence, the angle $\theta$ between the vector and a fixed direction in the transverse plane is
uniformly distributed in $[0,2\pi)$. Therefore, the pdf $P^c_v(\delta v_c)$ of
$\delta v_c$
is related to that of $\delta v$, $P_v(\delta v)$, by
\be \label{eq:project}
P^c_v(\delta v_c)=\frac{1}{\pi}\int_{|\delta v_c|}^{+\infty}
P_v(\delta v)\frac{d\delta v}{\sqrt{\delta v^2-\delta v^2_c}}.
\ee
Second, an ensemble of randomly oriented lines (with
uniform measure on a sphere, i.e. a uniform distribution of
initial solid angles $d\Omega_0$) will tend to concentrate along
directions of positive elongation. Thus, in order to compare model
results at later times with data that are taken at random
directions not correlated with the dynamics, the model results
need to be weighted with the evolving solid angle measure.
Conservation of fluid volume implies that $\ell^3 d\Omega_0 =
r(t)^3 d\Omega(t)$, i.e. in directions of growing $r(t)$, the
solid angle $d\Omega(t)$ decreases. Thus, probabilities must be
weighted by
\be
d\Omega(t)/d\Omega_0=[\ell/r(t)]^3.
\ee
Since $\dot{r}=\delta u\, r/\ell$, we can solve for $r(t)$ and
then obtain $d\Omega(t)/d\Omega_0 = \exp(-3\ell^{-1} \int_0^t
\delta u(t') dt')$. Using the solution for $\delta u$, we obtain
$d\Omega(t)/d\Omega_0=\ell^3[(\ell+\delta u_0 t)^2 + \delta v_0^2
t^2]^{-3/2}$ for $Q_0=0$. This factor is used to weight the
measured time-evolving pdfs from the model system. Note that when
$\delta v_0\to 0$ and $\delta u_0<0$, there is an unphysical
finite time singularity at $t\to\ell/|\delta u_0|$, when $r\to 0$.
\begin{figure}[ht]
\begin{minipage}[h]{0.65\linewidth}
\includegraphics[width=\linewidth]{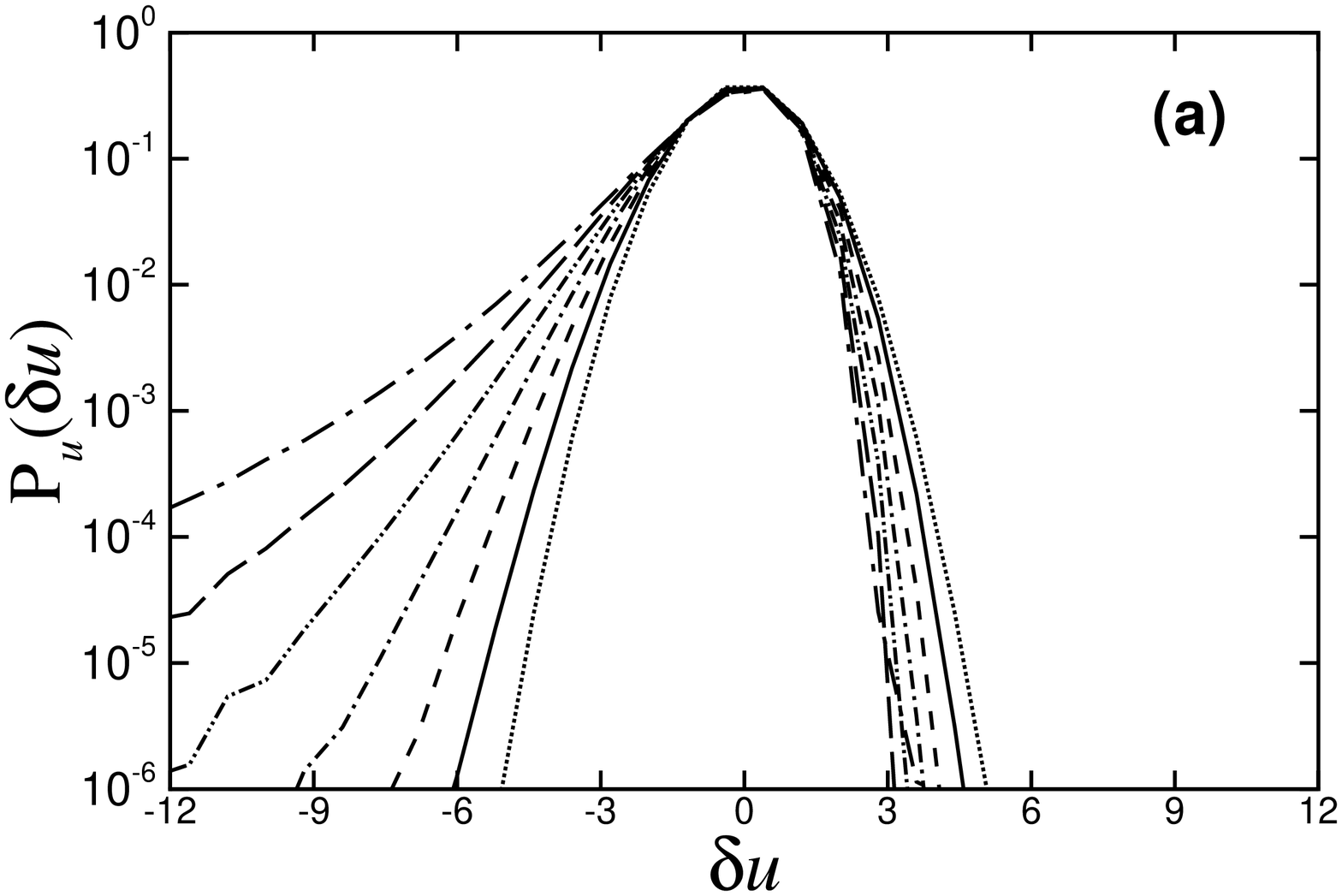}
\end{minipage}
\begin{minipage}[h]{0.65\linewidth}
\includegraphics[width=\linewidth]{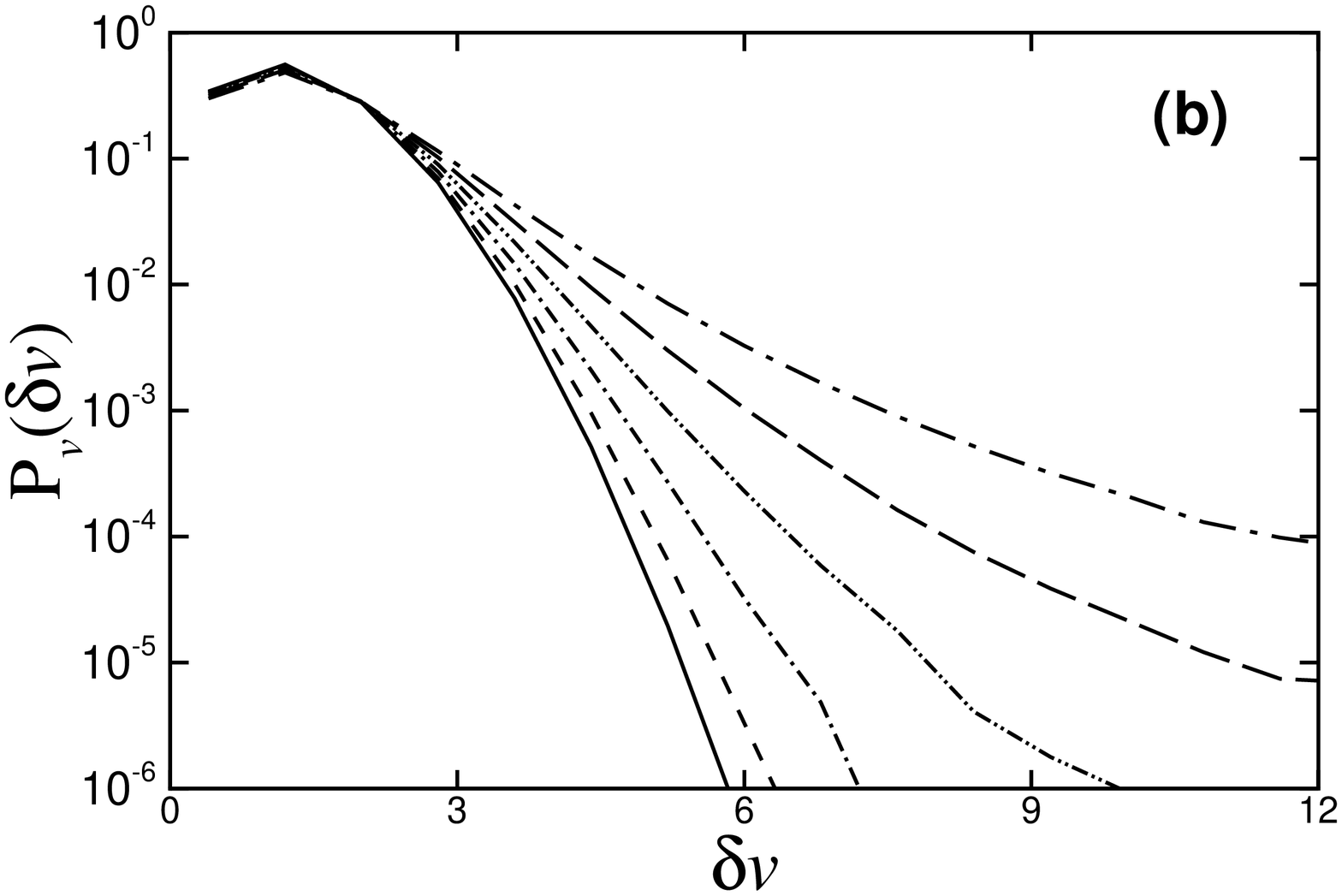}
\end{minipage}
\begin{minipage}[h]{0.65\linewidth}
\includegraphics[width=\linewidth]{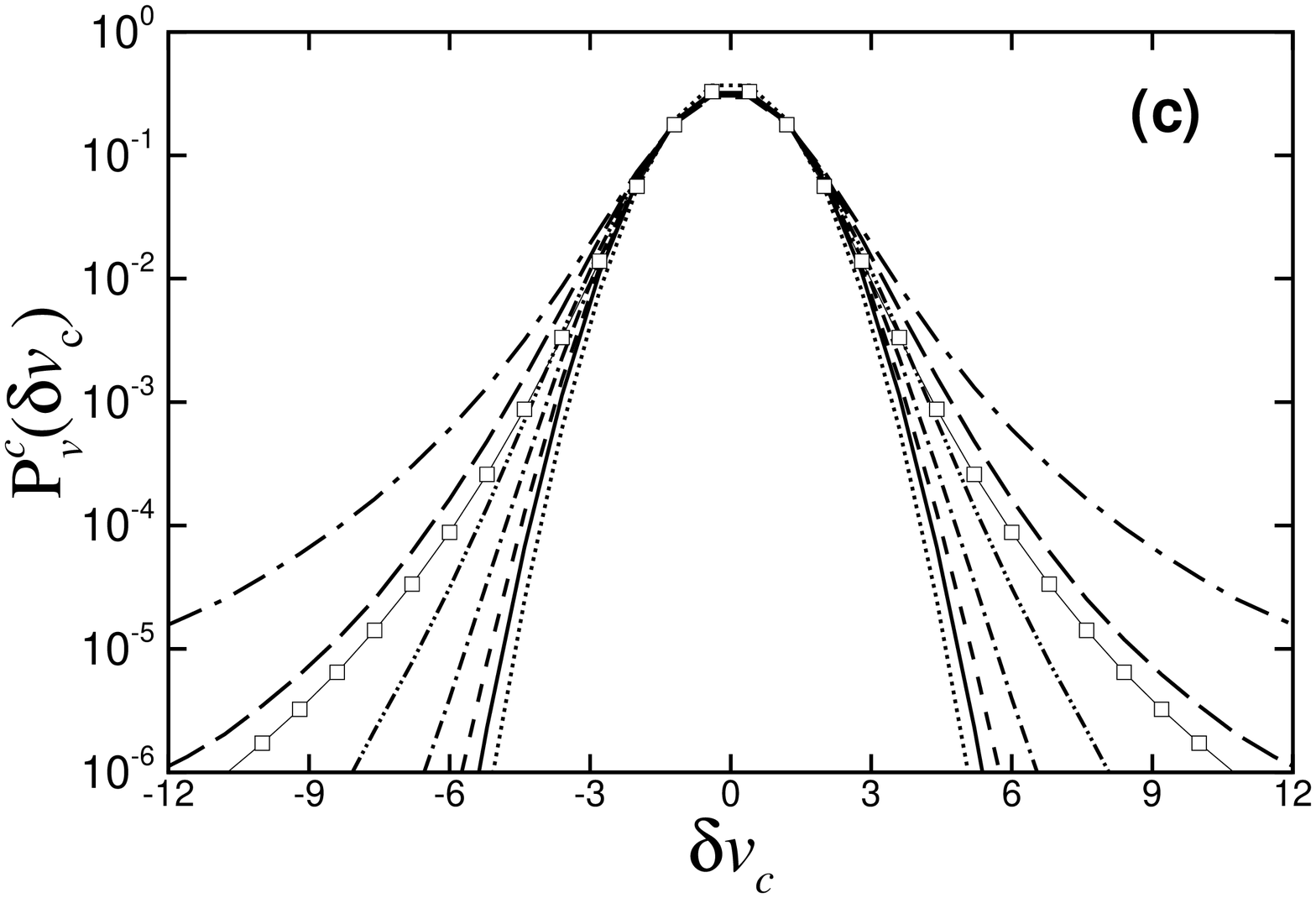}
\end{minipage}
\caption{Evolution of the pdf of velocity increments in time: (a)
longitudinal velocity increment, (b) the magnitude and (c) a
component of the transverse velocity increment vector. $\delta u$
is initialized as a standard Gaussian random number and $\delta
v$ as the square root of the sum of the square of two independent
Gaussian random numbers. (c) is calculated from (b) by numerical
integration of Eq. (\ref{eq:project}). For simplicity, we set
$\ell=1$, so that the characteristic time-scale of the ensemble is
$\tau = \ell/\delta u_0|_{\rm rms} = 1$. Dotted line in (a) and
(c): Gaussian; solid: $t=0.03$; dashed: $t=0.06$; dash-dotted:
$t=0.09$; dash-double-dotted: $t=0.12$; long-dashed: $t=0.15$;
long-dash-dotted: $t=0.18$; and thin line with squares in (c):
$t=0.18$ without correcting for evolving measure.}
\label{fig:pdf_uv}
\end{figure}

FIG. \ref{fig:pdf_uv} shows the evolution of the pdfs of the
longitudinal and transverse velocity increments (both the magnitude and a component), as time
progresses (for the case $Q_0=0$). It is immediately clear that
the two main qualitative trends observed in turbulence naturally
evolve from the solution of the  system: the skewness towards
negative values of longitudinal velocity increment, and the
noticeable flare-up of long tails in the pdfs of transverse
velocity increment. Also, these features appear rather quickly:
after a non-dimensional time $t/\tau = 0.18$ the pdf is already
highly skewed and displays stretched exponential tails. Very
similar results are observed for nonzero values of $Q_0$ (using
numerical forward time integration with a standard fourth order
Runge-Kutta routine, we tested $Q_0=\pm 2$): Relative to the
results for $Q_0=0$, the pdfs of $\delta u$ are shifted
to the left for $Q_0>0$ and to the right for $Q_0<0$, and only
very minor differences are seen for $\delta v$. The rapid
appearance of stretched exponential tails is due to the divergence
of the phase-space trajectories on the left half of the plane in FIG.
\ref{fig:phase}. For a given initial kinetic energy $\delta
u_0^2+\delta v_0^2$, if $\delta v_0$ is small, the invariant
$U_0$ can be arbitrarily large. Thus $\delta u$ and $\delta v$ can
later grow to very large values during the evolution.

In summary, the model system proves useful in showing that the
emergence of ubiquitous trends of 3D turbulence, namely
intermittent and asymmetric tails in pdfs of velocity increments,
occur even in the ``ballistic'' case ($Q_0=0$).  Considering all
possible random initial directions of relative motion, the
fraction of particle pairs that initially move towards each other
is small, thus large gradients in small spatial regions occur
rather infrequently but are very intense when they occur due to
the self-amplification mechanism for $\delta u$, and the
cross-amplification mechanism for $\delta v$.  While the model
system thus helps explain the origin and trends towards
intermittency in 3D turbulence, predicting quantitatively the
level of intermittency remains an open question. It requires
understanding the effects of pressure, inter-scale interactions
(that depends on interactions of vorticity and strains at various
scales, see e.g. \cite{Abidetal02,Taoetal02}) and viscosity that
are neglected in the model system. But already, the proposed model
system could be combined with cascade, mapping closure, or shell
models to enable these heuristic approaches to include a more
direct link to the underlying Navier-Stokes equations.

\begin{acknowledgments}
We thank Prof. Gregory Eyink for useful comments and for pointing out the need
to correct for the changing measure during the pdf evolution. We gratefully
acknowledge the support of the National Science Foundation (ITR-0428325 and CTS-0120317) and the
Office of Naval Research (N0014-03-0361).
\end{acknowledgments}


\end{document}